\documentclass[11pt]{article}   	
\usepackage{geometry}                		
\usepackage{graphicx}				
\usepackage{amssymb}
\usepackage{amsmath}
\usepackage{bm}
\usepackage{tikz}
\usetikzlibrary{decorations.pathmorphing,patterns,intersections,decorations.markings,arrows}
\newcommand\vecbf[1]{ {\bf #1} }
\newcommand\of[1]{\left( #1 \right)}
\newcommand\meskip{\, \, \, \, \, \, \, \, \, \, \, }

\newcommand\refeq[1]{(\ref{#1})}

\newcommand\dd{\partial}
\title{The (Weak) Gravitational Field of a Dirac Monopole}
\author{E.\ Banyas, J.\ Franklin \\ Physics Department, Reed College \\ jfrankli@reed.edu}
\date{}							

\begin{document}
\maketitle
\section{Abstract}

We establish the gravitational detectability of a Dirac monopole using a weak-field limit of general relativity, which can be developed from the Newtonian gravitational potential by including energy as a source. The resulting potential matches (by construction) the weak-field limit of two different solutions to Einstein's equations of general relativity: one associated with the magnetically monopolar spray of field lines emerging from the half-infinite solenoid that makes up the Dirac monopole, the other associated with the field-energetic source of the solenoid itself (the Dirac string).  The string's gravitational effect dominates, and we suggest that the primary strong-field contribution of the Dirac configuration is that of a half-infinite line of energy, whose GR solution is known.

\section{Introduction}

A Dirac monopole~\cite{DIRAC} is a magnetic configuration that, despite appearing monopolar, does not violate Maxwell's equation $\nabla\cdot\mathbf{B} = 0$.\footnote{There are, of course, other notions of magnetic monopoles (see~\cite{goddard, PRESKILL} for examples) but our work here is specific to the Dirac version.} To visualize the configuration, imagine an infinite solenoid of radius $R$ (which Dirac took to be zero) that has been cut in half, with one half removed.  The tail of the half-solenoid is out at spatial infinity, but the head has magnetic field lines spraying out of it, mimicking the field lines of an electric point charge. The setup is shown in Figure 1.

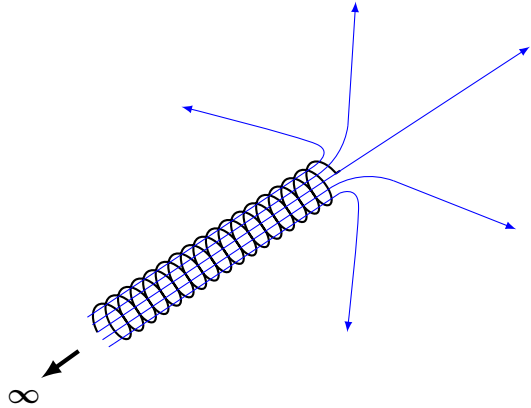
\begin{figure}[h!]\centering
\begin{tikzpicture}

\draw[thick,decorate,decoration={coil,amplitude=3mm,segment length=2mm}] (0,0) -- (3.225,2.15);

\draw[blue,->,>=latex]  (0,0) -- (3,2) -- (5.75,3.8);

\begin{scope}[yshift=-0.1cm,xshift=0.067cm]
\draw[blue]  (0,0) -- (3,2);
\draw[blue] (3,2) arc (125:70:0.95cm);
\draw[blue,->,>=latex] (3.87,2.115) -- (5.5,1.47);
\end{scope}

\begin{scope}[yshift=0.1cm,xshift=-0.067cm]
\draw[blue]  (0,0) -- (3,2);
\draw[blue] (3,2) arc (305:352:0.95cm);
\draw[blue,->,>=latex] (3.396,2.646) -- (3.50,4.3);
\end{scope}

\begin{scope}[yshift=-0.15cm,xshift=0.22cm]
\draw[blue] (0,0) -- (-0.07,-0.045);
\draw[blue]  (0,0) -- (3,2);
\draw[blue,->,>=latex] plot [smooth] coordinates {(3,2) (3.25,1.85) (3.10,0.15)};
\end{scope}

\begin{scope}[yshift=0.2cm,xshift=-0.134cm]
\draw[blue]  (0,0) -- (3,2);
\draw[blue,->,>=latex] plot [smooth] coordinates {(3,2) (3.05,2.30) (1.25,2.80)};
\end{scope}

\draw[ultra thick,->,>=latex] (-0.25,-0.25) -- (-0.75,-0.61);
\node at (-0.98,-0.85){$\bm{\infty}$};
\end{tikzpicture}
\caption{The Dirac monopole is the end of a half-infinite solenoid.}
\label{fig:Didea}
\end{figure}

In the $R \rightarrow 0$ limit, the magnetic field associated with the configuration, for a string that lies along the positive $\hat{\bf z}$ axis with opening at the origin, is (using the spherical coordinate $r \equiv \sqrt{x^2 + y^2 + z^2}$),
\begin{equation}\label{BFD}
{\bf B} = \frac{\mu_0 \, q_m}{4 \, \pi \, r^2} \, \hat{\bf r} - \mu_0 \, q_m \, \delta(x) \, \delta(y) \, \theta(z) \, \hat{\bf z},
\end{equation}
where $q_m$ is the effective pole strength.\footnote{In the Dirac model, there is no true monopole, just the monopole-\textit{like} spray of field from the end of the solenoid. For a true monopole (which would have $\nabla \cdot \vecbf B \ne 0$), $q_m$ would be the monopole's magnetic charge.} This magnetic field has zero divergence, thanks to the string contribution. We are taking the view that the string is ``real" (not just a manifestation of a gauge choice), and will probe the gravitational field that it (together with the monopolar piece) generates.

Dirac's motivation for introducing this model was to explain electric charge quantization.  The string plays a role similar to that of the infinite solenoid in the quantum mechanical Aharanov-Bohm effect.  By requiring that the phase of a wave function pick up an integer multiple of $2 \, \pi$ when going around the solenoid (rendering its presence undetectable), the relation between magnetic ``charge" ($q_m$) and electric charge $q$ must be
\begin{equation}\label{qqm}
\frac{\mu_0 \, q\, q_m}{4 \, \pi} =\frac{1}{2}\, n \, \hbar.
\end{equation}
Because the string is now electromagnetically undetectable, many authors attempt to remove it entirely,  either by chopping up space into disconnected pieces in which the string does not appear~\cite{WUYANG}, or by gauging it away as in~\cite{PRIMACK}.  But if we want to retain the usual notion of gauge transformation in E\&M as well as the condition that $\nabla \cdot \vecbf B = 0$, then the string cannot be a manifestation of a gauge choice, since its magnetic field appears in~\refeq{BFD}.

The outline of the paper is as follows:  We start by working in the static, linearized limit of general relativity, which can be thought of (here) as Newtonian gravity with energetic sources.  In that setting, we can describe a general solution for the gravitational contribution of any static electromagnetic configuration (in vacuum).  Using that general solution, and treating the string's magnetic field energy as a source, we can find the weak field gravitational potential associated with the magnetic field in~\refeq{BFD}.  From the linearized limit, we move on to the full gravitational field. If we think of the string as a solenoid of finite but small radius, then it is clear that the magnetic field, and hence the magnetic energy density inside the solenoid, must be large --- and infinite in the case that the radius goes to zero~\cite{PRESKILL} --- regardless of the size of $q_m$.  We can therefore approximate the space-time generated by the Dirac monopole as one generated by only a half-infinite line of source.  The metric in this case is known, and has the interesting property that for a particular value of energy density, the external metric is flat.  There is still a conical singularity at the source, of the usual topological sort~\cite{DESER}, but there is a notion of maximal undetectability for the string in this flat solution, which has no Newtonian analogue.  We explore this solution in the context of conical singularities, and contrast the Dirac string with the cosmic strings proposed in~\cite{Zeldovich} with weak field developed in~\cite{Vilenkin} and full relativistic metric from~\cite{Hiscock, Gott}.  In particular, we do not argue that there are many Dirac-type strings in the modern era --- unlike cosmic strings, there is no clear production mechanism.  Dirac proposed his strings without a concrete physical model, suggesting that they could exist just as monopoles might, produced early on, but remote and sparse.  Indeed, 
it only takes a single monopole to quantize electric charge, and there need be only one Dirac string attached to it.  Just as monopoles have yet to be discovered, their strings (should they exist) have also not been observed.  The point of the current work is to suggest an alternate method for detecting a Dirac monopole, and distinguishing it from the other theoretical monopoles that have been explored in the literature.

\section{Linearized Limit}

We can calculate the linearized gravitational field for the combination of terms in~\refeq{BFD}.  To be concrete, we will refer to the first term in~\refeq{BFD} as the ``monopole" contribution, and the second as the ``string" contribution.  We will calculate each one's gravitational field separately, and appeal to superposition (which exists in the linearized limit) to combine them.

The equation governing the potential $\phi$ in Newtonian gravity is the usual Poisson:
\begin{equation}\label{phipoisson}
\nabla^2 \phi = 4 \, \pi \, G \, \frac{u}{c^2},
\end{equation}
where $u$ is a static energy density.  We are interested in electric and magnetic field sources, so we will take the usual energy density from E\&M: $u = \frac{1}{2} \, \epsilon_0 \, E^2 + \frac{1}{2 \, \mu_0} \, B^2$.  In vacuum, both the electric and magnetic fields can be written in terms of scalar potentials, $\vecbf E = -\nabla V$ and $\vecbf B = -\nabla W$.  The solution to~\refeq{phipoisson} is
\begin{equation}\label{solution}
\phi = \frac{\pi \, G \, \epsilon_0}{c^2} \, \left( V^2 + c^2 \, W^2\right),
\end{equation}
since then:
\begin{equation}
\nabla^2 \phi = \frac{2 \, \pi \, G\, \epsilon_0}{c^2} \, \left(\nabla V \cdot \nabla V + c^2 \, \nabla W \cdot \nabla W + V \, \underbrace{\nabla^2 V}_{=0} + c^2 \, W \, \underbrace{\nabla^2 W}_{=0} \right) = 4 \, \pi \, G \, \frac{u}{c^2}.
\end{equation}

\subsection{Monopole Contribution}

Using the general solution in~\refeq{solution}, we can easily compute the monopole's gravitational contribution.  We know that the scalar magnetic potential for the monopole term is $W = \frac{\mu_0 \, q_m}{4 \, \pi \, r}$ and so we can extract the gravitational potential directly from~\refeq{solution}:
\begin{equation}
\phi_m = \frac{\mu_0 \, q_m^2\, G}{4 \, \pi \, c^2} \, \frac{1}{4 \, r^2}
\end{equation}
or, defining the ``length"\footnote{The dimensions of magnetic charge are ``electric charge times speed," so that $q_m/c$ is an electric charge.} $R_q \equiv \sqrt{\frac{\mu_0 \, q_m^2 \, G}{4 \, \pi \, c^4}}$, we have
\begin{equation}
\phi_m = \frac{R_q^2\, c^2}{4 \, r^2}.
\end{equation}
This potential matches the weak-field limit of the Reissner-Nordstr\"om metric, but with the electric charge replaced by a magnetic one, which is to be expected.

\subsection{The Half-Infinite Line of Source}

For the string portion of the model, we have a solenoid of radius $R$ with magnetic field $B_0$ inside it.  In order for the flux brought in by the solenoid to match the flux associated with the monopole, we must have $B_0 = \frac{\mu_0 \, q_m}{\pi \, R^2}$.  Then the energy density per unit length, $\lambda$, inside the solenoid is
\begin{equation}\label{l0string}
\lambda =\frac{1}{2 \, \mu_0} \, B_0^2 \, \pi \, R^2 = \frac{\mu_0\, q_m^2}{2\, \pi \, R^2}.
\end{equation}
Assuming the solenoid is basically a line source ($R \rightarrow 0)$\footnote{This is just to simplify notation; all we really need is to be away from the solenoid, so that the potential we find will be in vacuum, sourced by an infinite tube with constant $\lambda$ inside it.} with constant magnitude: 
\begin{equation}
u_s = \lambda \, \delta(x) \, \delta(y) \, \theta(z), 
\end{equation}
we can solve~\refeq{phipoisson} for this source, giving, in cylindrical coordinates (with $s \equiv \sqrt{x^2 + y^2}$),
\begin{equation}\label{stringerm}
\phi_s = -\frac{G \, \lambda}{c^2} \, \log\left( \frac{\ell \, \left(z + \sqrt{s^2 + z^2}\right)}{s^2} \right),
\end{equation}
where $\ell$ is an integration constant with the dimension of length.

\subsection{Full Potential}

Putting together the pieces, we have (in cylindrical coordinates now) a two-parameter ($R_q$, $\lambda$, associated with the size of the monopole charge $q_m$ and the radius of the solenoid respectively) family of solutions:
\begin{equation}\label{phiDS}
\phi_D = \phi_m + \phi_s = c^2 \, \left[ \frac{R_q^2}{4 \, (s^2 + z^2)} - \frac{G \, \lambda}{c^4} \,  \log\left( \frac{\ell \, \left(z + \sqrt{s^2 + z^2}\right)}{s^2} \right) \right].
\end{equation}
We have neglected the mass of the string, which is reasonable since we are focused on the electromagnetic source for gravity.  If we allowed the Dirac string to have some constant mass-per-unit-length $\bar\lambda$, the contribution to the potential would automatically combine with the magnetic $\lambda$, and provide no new terms, just a change in the effective source strength.

We can make some basic predictions about the type of motion experienced by a massive test particle moving under the influence of~\refeq{phiDS}.  For example, it is clear from the form of the potential that the $\hat{\bf z}$ component of angular momentum is conserved.  We will focus on orbital motion, and in particular, the pseudo-circular motion induced by the attractive solenoid source.  
Starting from the classical Lagrangian, in cylindrical coordinates, governing the motion of a particle of mass $m$ moving under the influence of $\phi_D$, 
\begin{equation}
L = \frac{1}{2} \, m \, \left[ \dot s^2 + s^2 \, \dot \phi^2 + \dot z^2 \right] - m \, c^2 \, \left[ \frac{R_q^2}{4 \, (s^2 + z^2)} - \frac{G \, \lambda}{c^4} \,  \log\left( \frac{\ell \, \left(z + \sqrt{s^2 + z^2}\right)}{s^2} \right) \right],
\end{equation}
(where dots refer to $t$-derivatives) we can set the constant $z$-component of angular momentum, $J_z = \frac{\dd L}{\dd \dot \phi} = m \, s^2 \, \dot \phi$, and write the Euler-Lagrange equations of motion for this Lagrangian as:
\begin{equation}
\begin{aligned}
\ddot s &= \frac{J_z^2}{m^2 \, s^3} + \frac{c^2 \, R_q^2 \, s}{2 \, \left( s^2 + z^2\right)^2} - \frac{G \, \lambda}{c^2} \, \left[ \frac{2 \, z}{s \, \sqrt{s^2 + z^2}} + \frac{s}{\sqrt{s^2 + z^2} \, \left( z + \sqrt{s^2 + z^2}\right)} \right] \\
\ddot z &= \frac{c^2 \, R_q^2 \, z}{2\, \left(s^2 + z^2 \right)^2} + \frac{G \, \lambda}{c^2} \, \left[ \frac{1}{\sqrt{s^2 + z^2 }} \right] \\
\dot\phi &= \frac{J_z}{m \, s^2}.
\end{aligned}
\end{equation}
If we introduce a constant $\ell_0$ with dimension of length, then we can define the dimensionless $S$ and $Z$ position variables: $s = \ell_0 \, S$ and $z = \ell_0 \, Z$.  Similarly, for a constant $t_0$ with dimension of time, we set the dimensionless $T$ via: $t = t_0 \, T$, and can write the equations of motion as (using primes to denote derivatives with respect to $T$)
\begin{equation}
\begin{aligned}
S''(T) &=  \frac{1}{S^3} +\frac{\alpha^2 \, S}{\left(S^2 + Z^2 \right)^2 } - \left[ \frac{2 \, Z}{S \, \sqrt{S^2 + Z^2}} + \frac{S}{\sqrt{S^2 + Z^2} \, \left( Z + \sqrt{S^2 + Z^2} \right)}  \right] \\
Z''(T) &= \frac{\alpha^2 \, Z}{\left(S^2 + Z^2\right)^2} +\ \frac{1}{\sqrt{S^2 + Z^2}} \\
\phi'(T) &= \frac{1}{S^2}
\end{aligned}
\end{equation}
by taking $t_0$ and $\ell_0$ so that $\frac{J_z \, t_0}{m \, \ell_0^2} = 1$ and $\frac{G \, \lambda \, t_0^2}{c^2 \, \ell_0^2} = 1$ and setting the dimensionless combination: $\alpha \equiv \frac{c \, R_q \, t_0}{\sqrt{2}\, \ell_0^2}$.

For the pure string portion of the motion, obtained by setting $\alpha = 0$, we can find pseudo-circular orbits in $S(T)$ (note that $Z(T)$ cannot be constant) by taking $S''(T) = 0$ for a given initial $Z(0) = Z_0$.  That defines an initially circular motion in the $x$-$y$ plane.  The value for $S(T)$ cannot be constant for all of the trajectory, since $Z(T)$ is not constant.  But for a relatively short period of time near $T = 0$, the motion is approximately circular for $\alpha = 0$.  Then, when we turn on $\alpha$, we expect the motion to lose its circular form.  For all cases, we must solve the equations of motion numerically, and we do that using a standard fourth-order Runge-Kutta ODE solver.

\begin{figure}[htbp] 
   \centering
  \includegraphics[width=4in]{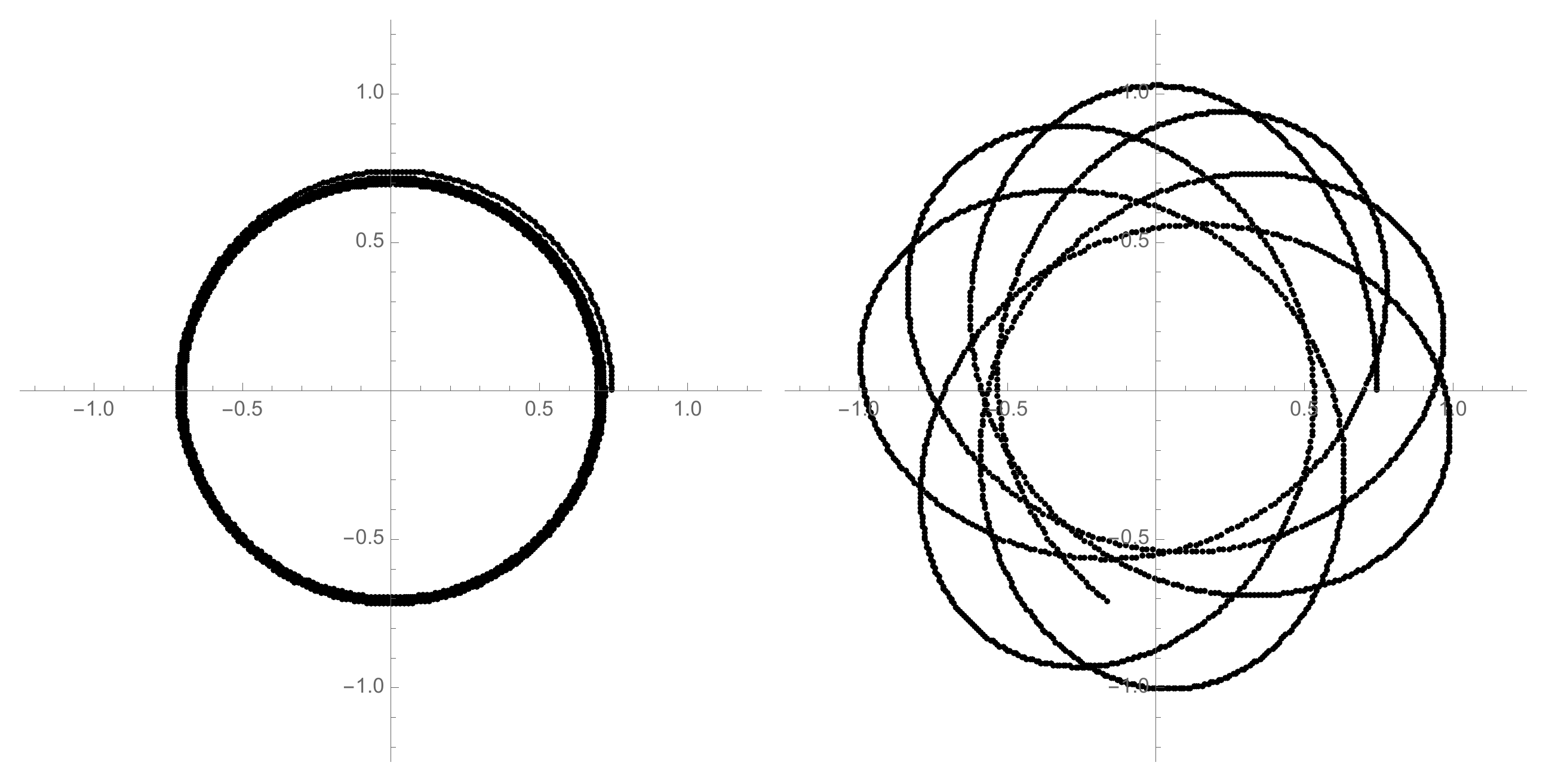} 
   \caption{The flattened trajectories (a view down the $\hat{\bf z}$ axis) of a test particle moving near the Dirac string.  The test particle starts from rest at $Z(0) = 1$, $S(0) \approx .745$ (to set an initially circular orbit).  On the left is the trajectory for $\alpha = 0$, and on the right, the trajectory for $\alpha = 2$.}
   \label{fig:traj}
\end{figure}

The particle tends to move upwards along the $z$ axis, which makes it difficult to compare the full three-dimensional trajectories.  To display the trajectories, we ``flatten" the motion by displaying only its projection into the $x$-$y$ plane.  In Figure~\ref{fig:traj}, we can see the $\alpha = 0$ motion on the left, with the motion for $\alpha = 2$ on the right.  We can clearly see the precessive effect of the monopole term.  The motion due to just the string and the string-plus-monopole are both relevant, since as we decrease the radius of the solenoid, that term's effect dominates, and we can capture this by setting $\alpha = 0$.  So the motion for $\alpha = 0$ is similar to what we expect in the zero-radius string case, or far away from the monopole.  The $\alpha = 2$ test particle motion is what we expect in the vicinity of the monopole.

\section{Strong Field Limit of the String}

Einstein's equations lack superposition, so there is no obvious way for us to generalize our weak field limit to a metric solution of GR.  However, 
referring to~\refeq{l0string}, we see that for any value of $q_m$, the energy density $\lambda$ will become arbitrarily large as the radius of the solenoid $R \rightarrow 0$.  The string portion of the source will therefore dominate the monopole term.  That source has cylindrical symmetry, and there is an infinite class of solutions to Einstein's equation associated with axially symmetric sources (the ``Weyl" metrics, see, for example~\cite{MTW}), including an infinite or half-infinite line of mass or energy.  Starting from the line element
\begin{equation}
ds^2 = -e^{2 \, a(s,z)} c^2 \, dt^2 + e^{2 \, \of{b(s,z) - a(s,z)}} \of{ds^2 + dz^2} + s^2 \, e^{-2 \, a(s,z)} d\phi^2,
\end{equation}
Einstein's equation in vacuum is satisfied by $a$ and $b$ with\footnote{All derivatives involved are the usual partial (not covariant) derivatives.}
\begin{equation}\label{bfroma}
\nabla^2 a = 0 \meskip \frac{\dd b}{\dd z} = 2 \, s \frac{\dd a}{\dd s} \, \frac{\dd a}{\dd z} \meskip \frac{\dd b}{\dd s} = s \, \left[ \of{\frac{\dd a}{\dd s}}^2 - \of{\frac{\dd a}{\dd z}}^2 \right].
\end{equation}
Taking $a = \phi_s/c^2$ (so that $a$ is dimensionless) from~\refeq{stringerm} (as $\nabla^2 \phi_s = 0$ away from the source, of course), and solving~\refeq{bfroma} for $b$,
\begin{equation}
b = -2 \, \of{\frac{G \, \lambda}{c^4}}^2 \log\of{\frac{s^2+ z^2+ z \, \sqrt{s^2 + z^2}}{s^2}}
\end{equation}
 gives us the well-known semi-infinite line of source solution~\cite{GAT} (see, for example~\cite{BONNOR}).  By construction, if we evaluate $g_{00}$ far away from the source, we will recover the appropriate linearized limit, with effective Newtonian potential $\phi_s$.

The metric associated with this exact solution has the interesting property that its Riemann curvature terms all have the form:
\begin{equation}
R^{\alpha}_{\, \, \beta \mu\nu} \sim \left( \frac{G \, \lambda}{c^4} - \frac{1}{2} \right) \hbox{$\times$ functions of $s$ and $z$}, 
\end{equation}
 so that if the magnetic field has energy density 
\begin{equation}
\lambda = \frac{c^4}{2\, G},
\end{equation}
the space-time is flat away from the line of source~\cite{BONNOR, BG}.  This anomaly has no weak-field analogue; there is no value for $\lambda$ (other than $\lambda = 0$) that makes the Newtonian gravitational force disappear.  While this value for the energy density leads to no space-time curvature, there remains a conical singularity~\cite{BG}, so presumably the string is still detectable (via the ``focusing trajectories" for test particles described in~\cite{BG} and in the context of light deflection, discussed below).  Nevertheless, at this special value for $\lambda$, and assuming we take $n = 1$ in~\refeq{qqm}, we can find the radius of the solenoid that carries the magnetic field: $R \approx 2 \times 10^{-34}$ m (one order of magnitude larger than the Planck length, where presumably, additional quantum mechanical physics plays a role).

\section{Comparison with Cosmic Strings}

Cosmic strings represent one of three topological structures that can arise due to spontaneous symmetry breaking~\cite{Kibble,hindmarsh,Vilenkin} of a scalar field.  They could have been formed as the universe cooled, and can persist in a stable manner.  In the simplest model, the strings are infinitely long, straight lines of energy-density source that can have finite radius or not.  In that sense, they are similar to the half-infinite line of energy-density source that we have considered here.  However, there is an important difference:  The source structure of cosmic strings, even ones with zero radius, is informed by the stress tensor of the scalar fields that generated them $T^{\mu\nu} \sim \phi_,^{\, \, \mu}  \phi_,^{\, \, \nu} - 1/2 \eta^{\mu\nu}  \phi_,^{\, \, \alpha} \phi_{, \alpha}$.  For a source lying along the $z$-axis, the stress tensor (density) has $T^{0}_{\, \, 0} = T^{z}_{\, \, z}$ with all other components zero.\footnote{For a zero-radius, infinitely long source lying along the $z$-axis, the stress tensor density only has coordinate dependence on $x$ and $y$.  The $T^x_{\, \, x}$ and $T^{y}_{\, \, y}$ components vanish by conservation of $T^{\mu\nu}$.}  This means, in particular, that the trace of the stress tensor is non-zero.  For the Dirac string, the source is electromagnetic, and must be informed by the vector stress tensor of E\&M, for which the trace $T = 0$.   That difference in source makes a difference in the gravitational field.  In particular, Einstein's equation tells us that for $T = 0$, the scalar curvature $R = 0$ as well, and so it is just the Ricci tensor (and not the full Einstein tensor) that is sourced by the stress tensor.

In the cosmic string case, the presence of a non-trace-free stress tensor allows the $T^z_{\, \, z}$ component to act as a source in each of the weak-field equations,\footnote{Writing Einstein's equation in terms of the trace $T$ instead of the scalar $R$, we have
\begin{equation*}
R^{\mu\nu} = \frac{8 \pi G}{c^4} \of{ T^{\mu\nu} - \frac{1}{2} \eta^{\mu\nu} T},
\end{equation*}
and the trace appears in each field equation.  The situation persists for the weak-field case, see~\cite{Vilenkin}, for example.} and similarly in the full case considered by~\cite{Hiscock} and~\cite{Gott}.  In particular, one can show that the $g_{00}$ component of the metric has no net source for vacuum cosmic strings, and the metric reduces to a flat metric with a conical singularity, manifest as an angular coordinate that goes from $\phi = 0$ to $\of{1 - 4 G \lambda/c^4}2 \pi$ where $\lambda$ is the energy-per-unit-length of the cosmic string.  This is the weak-field result of~\cite{Vilenkin}. The full general relativistic result of~\cite{Hiscock, Gott}, in which cosmic strings with finite radius are considered,\footnote{One could also perform that interior analysis for a finite-radius Dirac string.  That would be interesting, but would not change the external field significantly (beyond connecting its constants more explicitly to the interior), nor would it prevent charge quantization.} has external vacuum that is also conical, and with angular deficit that agrees with the weak-field limit for $\lambda$ small.  In all of these, the source mass provides the angular deficit.  One can use that angular deficit to find the deflection of light by the cosmic string, 
\begin{equation}
\delta \phi = \frac{4 \pi G \lambda}{c^4}.
\end{equation}

The Dirac string, with its electromagnetic stress tensor source, does not have flat solutions with conical singularities for all values of the energy density.  At the critical value we found above, the conical singularity should have light-deflection angle that goes like $G \lambda/c^4$ as in the cosmic string case, but for $\lambda = c^4/(2 G)$, the deflection angle is $\sim G\lambda/c^4 \sim 1$, which is large, and would be obviously detectable by lensing if there were Dirac strings around in any abundance.  Interestingly, then, the single value that would render the Dirac string locally undetectable (by producing a  flat space-time) would lead to detection through a different mechanism.  Of course, if Dirac strings did exist, their monopolar heads would also be observable electromagnetically, and those are not in evidence.

\section{Conclusion}

While the string portion of the Dirac monopole is not detectable via an electromagnetic interaction, it {\it is} detectable gravitationally.  The energy density of the monopole's magnetic field sets up a gravitational field that extends to the region outside the string.  There are two contributions to that field: the magnetic field in the solenoid itself, and the magnetic field lines that spray out of the end (which we have modeled as a point magnetic monopole).  The Newtonian potential has limiting cases $\lambda = 0$ and $R_q=0$, that correctly match the weak-field limit of the point magnetic monopole (a Reissner-Nordstr\"om-type metric~\cite{MTW}) and a half-infinite line of energy density (a Weyl-type metric) individually.  We have explored some of the interesting trajectories for a test particle moving under the influence of the Newtonian potential to ensure that they are not trivial, and are detectable in principle.

We do not know the exact general relativistic solution to our line-plus-point source model, but it is clear from the relation between the magnetic charge and the constant magnetic field inside the solenoid from~\refeq{l0string} that in the limit $R\rightarrow 0$, the half-infinite portion of the solution will dominate.  So, in a sense, the Dirac ``monopole" has a gravitational field that is just that of the string itself, with the magnetic monopole's contribution negligible.  In that limiting case, we {\it do} know the full general relativistic solution.

There are interesting features of the general relativistic half-infinite line of source solution that might be relevant to the Dirac string.  For example, there is an anomalously flat value for $\lambda$ --- if $\lambda = c^4/(2 \, G)$, the space-time surrounding the string has vanishing Riemann tensor, and hence represents a flat solution. If one demanded that the Dirac string be almost entirely physically undetectable, then it is tempting to set the energy density, and hence the solenoid radius $R$, to achieve this special value of $\lambda$. That would provide a prediction about the allowed size of the Dirac string, one which is otherwise missing.  However, even this flat solution would be detectable by conical lensing of light.  In the end, we do not imagine that the universe is filled with Dirac strings, produced by some mechanism early on.  The point here is, instead, to establish that if Dirac strings exist, they would be detectable gravitationally, and should be considered observable.

\section{Acknowledgements}
We thank Professor Griffiths for helpful commentary, and for suggesting the solution in~\refeq{solution} and Professor Deser, who provided monopolar insight from the start of this project.  We have benefitted from the thoughtful suggestions of our referees.

\end{document}